\documentclass[conference,letterpaper,10pt]{IEEEtran}
\IEEEoverridecommandlockouts

\usepackage[T1]{fontenc}
\usepackage[english]{babel}
\usepackage{textcomp}
\usepackage[table, pdftex, usenames, dvipsnames]{xcolor}
\usepackage{xspace}
\usepackage{graphicx}
\usepackage{amsfonts}
\usepackage{amsmath}
\usepackage{amssymb}
\usepackage{booktabs}
\usepackage[acronym]{glossaries}
\usepackage[capitalize,noabbrev]{cleveref} 
\usepackage{cite}
\usepackage{tikz}
\usepackage{multirow}
\usepackage{multicol}
\usepackage{colortbl}
\usepackage{subfigure}
\usepackage{siunitx}
\usepackage{algpseudocode}
\usepackage{algorithm}
\usepackage{url}
\usepackage[utf8]{inputenc}
\usepackage[most]{tcolorbox}

\glsdisablehyper
\AtBeginDocument{}
\tcbset{
    colback=gray!5,
    colframe=gray!40,
    left=6pt,
    right=6pt,
    top=6pt,
    bottom=6pt,
    boxsep=0pt,
}

\makeatletter
\DeclareRobustCommand\onedot{\futurelet\@let@token\@onedot}
\def\@onedot{\ifx\@let@token.\else.\null\fi\xspace}
\def\eg{\emph{e.g}\onedot} 
\def\ie{\emph{i.e}\onedot}

\makeatother

\begin{document}

\title{
\begin{tikzpicture}[overlay, remember picture]
    \node at ([yshift=-1.3cm]current page.north) {
        \normalsize\textcolor{gray}{This paper has been accepted for publication at the }
    };
    \node at ([yshift=-1.8cm]current page.north) {
        \normalsize\textcolor{gray}{IEEE International Instrumentation and Measurement Technology Conference (I2MTC), Nancy, France, 2026}
    };
\end{tikzpicture}

Performance Analysis of Edge and In-Sensor AI Processors: A Comparative Review
\thanks{This work was funded by the Swiss National Science Foundation (Grant 219943).
\textsuperscript{\textsection}L. Capogrosso and P. Bonazzi contributed equally to this work.}
}

\author{
\IEEEauthorblockN{
Luigi Capogrosso\IEEEauthorrefmark{1}\textsuperscript{\textsection},
Pietro Bonazzi\IEEEauthorrefmark{2}\textsuperscript{\textsection},
Michele Magno\IEEEauthorrefmark{2}\IEEEauthorrefmark{1}}
\IEEEauthorblockA{
\IEEEauthorrefmark{1}Interdisciplinary Transformation University of Austria,
\IEEEauthorrefmark{2}ETH Zurich}
}
\maketitle

\bstctlcite{IEEEexample:BSTcontrol}

\begin{abstract}
This review examines the rapidly evolving landscape of ultra-low-power edge processors, covering heterogeneous Systems-on-Chips (SoCs), neural accelerators, near-sensor and in-sensor architectures, and emerging dataflow and memory-centric designs.
We categorize commercially available and research-grade platforms according to their compute paradigms, power envelopes, and memory hierarchies, and analyze their suitability for always-on and latency-critical Artificial Intelligence (AI) workloads.
To complement the architectural overview with empirical evidence, we benchmark a 336 million Multiply–Accumulate (MAC) segmentation model (PicoSAM2) on three representative processors: GAP9, leveraging a multi-core RISC-V architecture augmented with hardware accelerators; the STM32N6, which pairs an advanced ARM Cortex-M55 core with a dedicated neural architecture accelerator; and the Sony IMX500, representing in-sensor stacked-Complementary Metal-Oxide-Semiconductor (CMOS) compute.
Collectively, these platforms span MCU-class, embedded neural accelerator, and in-sensor paradigms.
The evaluation reports latency, inference efficiency, energy efficiency, and energy–delay product.
The results show a clear divergence in hardware behavior, with the IMX500 achieving the highest utilization (86.2 MAC/cycle) and the lowest energy–delay product, highlighting the growing significance and technological maturity of in-sensor processing.
GAP9 offers the best energy efficiency within microcontroller-class power budgets, and the STM32N6 provides the lowest raw latency at a significantly higher energy cost.
Together, the review and benchmarks provide a unified view of the current design directions and practical trade-offs that are shaping the next generation of ultra-low-power and in-sensor AI processors.
\end{abstract}

\begin{IEEEkeywords}
Benchmarking, Edge AI, In-Sensor Computing, Ultra-Low-Power Processors.
\end{IEEEkeywords}

\section{Introduction} \label{sec:intro}

Edge AI is rapidly moving computation from cloud servers to devices operating close to where data is generated.
This shift is driven by the need for low latency, reduced communication overhead, improved privacy, and continuous context-aware processing \cite{Wang2025}.
These requirements span wearables, Internet of Things (IoT) nodes, wireless sensor networks, autonomous micro-robots, and Augmented Reality (AR)/Virtual Reality (VR) systems \cite{Singh2023}, all of which must run Machine Learning (ML) workloads under stringent energy and thermal constraints \cite{Mao2024}.

Most edge systems operate below a practical power ceiling of roughly 200 mW \cite{Fanariotis2023}.
This limit arises not only from battery capacity but also from thermal dissipation, form-factor constraints, and the need to keep sensing, communication, and inference active without dedicated cooling or high-power regulators \cite{Giordano2020}.
Meanwhile, Deep Neural Networks (DNNs) have grown in size and computational demand \cite{Thompson2020} much faster than process technology can compensate \cite{Nechi2023}.
As a result, efficient on-device inference has become the dominant bottleneck in modern edge systems \cite{Zhao2024}.
This challenge is amplified by the fact that energy is not consumed by computation alone: sensing, signal conditioning, memory access, and wireless transmission all draw from the same constrained power budget.
In battery-operated or thermally limited devices, these components must operate continuously and cooperatively within a tight energy envelope, making holistic system-level efficiency as critical as algorithmic or architectural optimization \cite{Bougioukou2025}.

\begin{figure*}[t!]
    \centering
    \includegraphics[width=.95\linewidth]{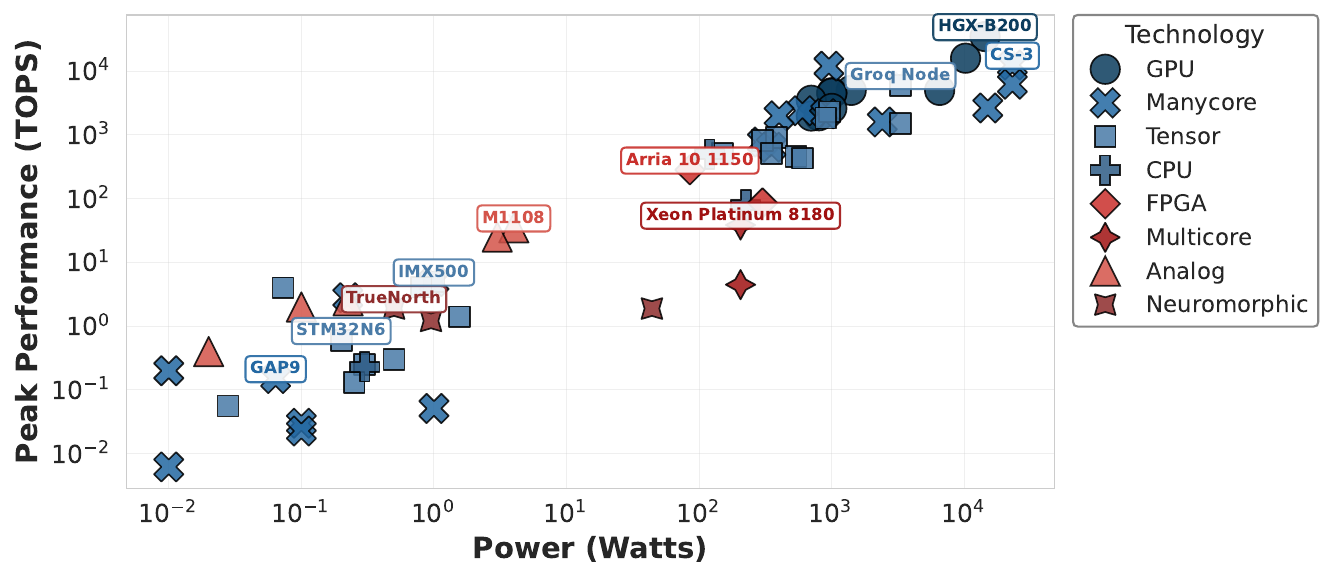}
    \caption{Peak performance in TOPS vs. power consumption of publicly announced AI accelerators and processors.
    Data are from \cite{Reuther2025, STM32N6, IMX500, RenesasRA}.}
    \label{fig:top10_tech}
\end{figure*}

These constraints have driven the development of specialized processors optimized for low-power Artificial Intelligence (AI) \cite{Dhilleswararao2022}, to maximize operations per watt and deliver efficient inference under strict energy budgets \cite{Sun2024, Baek2025}.
They exemplify the architectural trade-offs available to designers: high clock speed versus power consumption, local memory size versus data movement energy, and dedicated Multiply–Accumulate (MAC) arrays versus general-purpose programmability.

Edge AI workloads can no longer rely on pre-AI-generation Microcontroller Unit (MCU) cores such as ARM Cortex-M4 or Cortex-M33 \cite{LucanOrașan2022}, whose energy and throughput limitations make them unsuitable for sustained on-device inference \cite{Millar2025}.
This gap has catalyzed the emergence of domain-specific architectures, such as neural accelerators, dataflow engines, RISC-V vector and custom extensions, compute-in-memory arrays, and in-sensor processors, which are explicitly optimized for high compute density (Tera-Operations Per Second (TOPS)/W), minimal data movement, and ultra-low-power operation \cite{Silvano2025}.
Maximizing operations per watt has become the dominant design objective, a trend reflected in recent commercial and research platforms: the Google Edge Tensor Processing Unit (TPU) and Intel Movidius Myriad prioritize TOPS/W, the PULP/GAP9 processor targets maximum performance within IoT-class energy envelopes, novel ARM Cortex-M55 and Cortex-M85 cores with ARM Ethos-U55/U65 accelerators bring ML-efficient datapaths into MCU-scale devices, and Sony's IMX500 integrates in-sensor compute to drastically reduce data transfer overhead \cite{ECS2025}.
These developments span a broad continuum of architectural strategies, ranging from neuromorphic and compute-in-memory approaches that break von Neumann constraints \cite{Aguirre2024}, to increasingly capable ARM Cortex-M/A and customizable RISC-V platforms, and finally to fully integrated in-sensor processors and higher-power edge Neural Process Units (NPUs).
Together, they illustrate how the field has diversified to meet the stringent power, latency, and form-factor constraints that characterize modern Edge AI systems.
Some notable examples include sub-threshold MCUs (\eg{}, Ambiq \cite{Apollo5}), smart-sensing System-on-Chips (SoCs) (\eg{}, Sony CXD5602 \cite{CXD5602}), heterogeneous ARM Cortex-M55 MCUs with NPUs (\eg{}, STM32N6 \cite{STM32N6}), and multi-core accelerators such as GreenWaves GAP9 \cite{Mueller2024}.

Beyond reviewing these architectures, this work provides a direct and quantitative comparison of three heterogeneous Edge AI platforms, \ie{}, GAP9, STM32N6, and Sony IMX500 \cite{IMX500}, running the same lightweight PicoSAM2 segmentation model (336 million MAC) \cite{Bonazzi2025}; see \cref{fig:comparison_plots}.
Using cycle-accurate profiling and power measurements, we evaluate each system using four hardware-centric metrics: latency per inference, compute utilization (MAC/cycle), energy efficiency (MAC/J), and Energy–Delay Product (EDP).
This controlled evaluation reveals the practical limits, strengths, and bottlenecks of each architecture: the IMX500 achieves the highest compute density and the lowest EDP; GAP9 delivers competitive MAC/J despite lower frequency; and STM32N6 provides the shortest latency at substantially higher power.

By combining a survey of emerging sub-200-mW architectures with a grounded, model-matched benchmark across three representative platforms, this work aims to clarify the technology landscape and expose the design trade-offs that define AI computing at the edge.
To visualize this context, \cref{fig:top10_tech} maps the peak performance versus power consumption of current state-of-the-art accelerators \cite{Reuther2025, STM32N6, IMX500, RenesasRA}, highlighting the distinct operating region of ultra-low-power devices.
\section{Background} \label{sec:background}

\subsection{Edge Computing and Edge AI Tasks}
Edge systems interface directly with the physical world via mixed-signal front ends that acquire, digitize, and buffer sensor data.
Traditional architectures rely on high-fidelity Analog-to-Digital Conversion (ADC), but emerging in-sensor computing platforms increasingly embed intelligence within the sensing pipeline to reduce data movement at the source \cite{Fabre2024}.

Inference at the edge executes trained models in real time under tight thermal and energy constraints—often below 10 mW in in-sensor scenarios.
Full-precision execution is typically prohibitive, making hardware-aware optimizations a core design requirement \cite{Capogrosso2024}.
Common compression and optimization methods include quantization, pruning, and knowledge distillation.

\subsection{Energy Consumption Components}
The energy profile of an edge node is dominated by sensing, computation, and communication \cite{Katare2023}.
The sensing chain, \ie{} Analog Front-End (AFE) + ADC, is negligible for low-bandwidth modalities, but scales strongly with sampling rate and bit depth for vision or high-frequency signal domains.
Many systems, therefore, rely on low-power trigger paths to keep high-fidelity ADCs dormant except during relevant events.

The computational energy depends on the model’s arithmetic intensity and the frequency/voltage settings of the Central Processing Unit (CPU), Digital Signal Processing (DSP), or NPU.
Communication typically dwarfs both: wireless links consume nanojoules to microjoules per bit, whereas a modern MAC operation costs only picojoules.
This disparity explains why local inference, even at non-zero cost, reduces overall system energy by minimizing radio activity.

\subsection{Metrics to Evaluate Low-Power AI Platforms}
The efficiency of an edge processor is determined by both its computational behavior and its energy usage for a fixed model workload.
In this work, we rely on a set of hardware-level metrics that capture these dimensions without ambiguity.

\emph{\textbf{Latency per Inference (ms).}}
Latency is the end-to-end time required to execute a single forward pass of the model.
It reflects the combined influence of clock speed, memory hierarchy, and accelerator utilization.

\emph{\textbf{Inference Efficiency (MAC/cycle).}}
MAC/cycle measures how effectively the architecture sustains its theoretical compute throughput, capturing utilization, dataflow efficiency, and the impact of memory stalls.
Higher values indicate that the accelerator or DSP is well matched to the model structure.

\emph{\textbf{Energy Efficiency (MAC/J).}}
MAC/J quantifies the number of useful operations performed per joule.
It consolidates dynamic power, leakage, and memory movement into a single figure of merit.
Systems with optimized datapaths, short memory routes, or reduced-precision execution typically achieve higher MAC/J.

\emph{\textbf{Energy–Delay Product (EDP).}}
EDP combines the energy consumed per inference with its corresponding latency.
The metric penalizes architectures that are fast but power-hungry, as well as those that are energy-efficient but slow.
A lower EDP indicates a more balanced platform that offers both low latency and low energy cost.

\emph{\textbf{Energy per Inference (µJ or mJ).}}
This metric quantifies the total energy required to process a single input sample through the full model.
In ultra-low-power applications, minimizing the energy per inference is often prioritized over throughput because it directly affects the maximum number of events the battery can sustain.

\emph{\textbf{Peak Power and Average Power.}}
Peak power corresponds to the maximum instantaneous current the hardware can draw, a critical constraint for power sources with high internal impedance, such as coin-cell batteries or energy harvesters.
The average power represents the entire duty cycle—including active and sleep periods—and is used to size the battery capacity (mWh) needed to meet the operational lifetime target (\eg{}, 5–10 years).

\emph{\textbf{Memory Footprint.}}
Memory requirements are typically divided into non-volatile memory (Flash) for model weights and volatile memory (SRAM) for activation maps and input buffers during execution. 
\section{Taxonomy of Low-Power Edge Processors} \label{sec:taxonomy}

\begin{table*}[t!]
    \centering
    \caption{Hardware performance comparison for devices under 2 W (sorted by Efficiency: TOPS/W).
    Data from \cite{Reuther2025, STM32N6, IMX500, RenesasRA}.}
    \begin{tabular}{l c c c c c}
\toprule
\textbf{HW Platform} & \textbf{Peak Perf. (TOPS)} & \textbf{Power (W)} & \textbf{Precision} & \textbf{HW Architecture} & \textbf{Efficiency (TOPS/W)} \\
\midrule
Netcast                      & 1.00E+01  & 0.001 & int8  & Dataflow ASIC              & 1.00E+04 \\
Ergo                         & 4.00E+00  & 0.073 & int8  & Tensor ASIC                & 5.48E+01 \\
Ethos N77                    & 4.10E+00  & 0.800 & int8  & Tensor ASIC                & 5.13E+00 \\
MX3                          & 5.00E+00  & 1.000 & fp16  & Manycore ASIC              & 5.00E+00 \\
Tianjic                      & 1.21E+00  & 0.950 & int8  & Neuromorphic               & 1.27E+00 \\
AML200                       & 2.00E+00  & 0.100 & int8  & Analog In-Memory           & 2.00E+01 \\
GAP9                         & 1.51E-01  & 0.0640 & int8  & RISC-V Manycore            & 2.36E+00 \\
AIStorm                      & 2.50E+00  & 0.225 & int8  & Analog Compute-in-Sensor   & 1.11E+01 \\
Gyrfalcon                    & 2.80E+00  & 0.224 & int8  & Manycore ASIC              & 1.25E+01 \\
AML100                       & 4.00E-01  & 0.020 & int8  & Analog In-Memory           & 2.00E+01 \\
STM32N6                      & 6.00E-01  & 0.200 & int8  & ARM Cortex-M55 + NPU       & 3.00E+00 \\
Cortex-M85 (STM32V8/RA8)     & 1.30E-01  & 0.250 & int8  & ARM Cortex-M85             & 5.20E-01 \\
NDP101                       & 2.00E-01  & 0.010 & int4  & RISC-V + HW Acc            & 2.00E+01 \\
NDP200                       & 6.20E-03  & 0.010 & int8  & RISC-V + HW Acc            & 6.20E-01 \\
NDP250                       & 3.00E-02  & 0.100 & int8  & RISC-V + HW Acc            & 3.00E-01 \\
IMX500                       & 7.952E-02 & 0.016 & int8  & Manycore ASIC              & 4.97E+00 \\
Max 78000                    & 5.60E-02  & 0.028 & int8  & Tensor Accelerator MCU     & 2.00E+00 \\
GAP8                         & 2.27E-02  & 0.100 & int8  & RISC-V Manycore            & 2.27E-01 \\
Eyeriss                      & 6.72E-02  & 0.278 & int16 & Dataflow ASIC              & 2.42E-01 \\
ShiDianNao                   & 1.94E-01  & 0.320 & int16 & Dataflow ASIC              & 6.06E-01 \\
DianNao                      & 4.52E-01  & 0.485 & int16 & Dataflow ASIC              & 9.32E-01 \\
PuDianNao                    & 1.06E+00  & 0.596 & int16 & Dataflow ASIC              & 1.78E+00 \\
EIE                          & 1.02E-01  & 0.600 & int16 & Dataflow ASIC (Sparse)     & 1.70E-01 \\
K210                         & 2.50E-01  & 0.300 & int8  & RISC-V Dual Core + KPU     & 8.33E-01 \\
Kendrite K210                & 2.30E-01  & 0.300 & int8  & RISC-V Dual Core + KPU     & 7.67E-01 \\
TrueNorth                    & 1.89E+00  & 0.500 & int8  & Neuromorphic               & 3.78E+00 \\
KL520 NPU                    & 3.00E-01  & 0.500 & int8  & Tensor ASIC                & 6.00E-01 \\
xcore.ai                     & 5.12E-02  & 1.000 & int8  & DSP-like Multicore         & 5.12E-02 \\
KL720                        & 1.40E+00  & 1.556 & int8  & Tensor ASIC                & 9.00E-01 \\
\bottomrule
\end{tabular}
    \label{tab:hw_low_power_efficiency}
\end{table*}

\subsection{MCU Processors with ML Extensions}
Modern MCUs for Edge AI rely on sub-threshold voltage operation, aggressive clock gating, and DSP or low-power ML extensions to balance throughput with tight energy budgets.
Limited on-chip SRAM/Flash resources drive efficient memory hierarchies and quantization.
Ambiq Apollo510 \cite{Apollo5}, built around the ARM Cortex-M55, uses Subthreshold Power Optimized Technology (SPOT) technology for extremely low active power while maintaining real-time inference \cite{Schilk2023}.
STM32U5/H5 \cite{STM32U5, STM32H5} utilizes Arm Cortex-M33 cores with advanced power modes (U5) or higher-frequency computing (H5), both of which integrate TrustZone and math accelerators \cite{Bartoli2025}.
Nordic’s nRF54 series \cite{nRF54L} uses a 22 nm multicore ARM Cortex-M33 architecture to separate the radio and application domains, thus increasing performance per watt.
The Renesas RA \cite{RenesasRA} and Infineon DA \cite{InfineonDA} series, which typically integrate Cortex-M33, M55, or M85 cores, extend secure, low-power MCU designs with expanded memory and peripheral integration.

\subsection{MCU with Dedicated ML Accelerators}
These devices pair an MCU with Convolutional Neural Network (ConvNet) engines or NPUs, using Single Instruction Multiple Data (SIMD) or MAC arrays and INT8 or binary quantization, for high-throughput inference. 
MAX78000 \cite{MAX78000} couples a ConvNet accelerator with Cortex-M4F and a RISC-V coprocessor, using on-chip weight/data memory to avoid external fetches \cite{Moss2022}.
The Sony CXD5602 \cite{CXD5602} utilizes a hex-core Cortex-M4F design on the Fully Depleted Silicon-On-Insulator (FD-SOI) for efficient parallel processing of sensors \cite{Chakraborty2025} and Global Navigation Satellite System (GNSS).
GAP9 \cite{Mueller2024} provides a scalable RISC-V cluster with configurable precision and 1.6 MB of L2 SRAM for software-defined ML acceleration, currently used mainly in vision \cite{Niculescu2025} and nano-drone applications \cite{Crupi2025}.
NXP i.MX RT1170 \cite{iMXRT1170} combines a 1 GHz M7 with an M4 and 2 MB SRAM to bridge MCU–application processor performance.
Kendryte K210/K230 \cite{K210, K230} introduces RISC-V SoCs with Kernel Processing Unit (KPU) acceleration, with the K230 supporting external Low-Power Double Data Rate (LPDDR) and DDR for larger models.
STM32N6/STM32V8 \cite{STM32N6, STM32V8} integrates Neural-ART NPU (600 Giga-Operations Per Second (GOPS)) and the new Cortex-M85 with Helium for high-performance AI and signal processing.
In particular, the V8 is among the first commercial MCU-class processors to combine this advanced core with an Ethos-U65 NPU, providing a tightly coupled AI acceleration pipeline for embedded workloads \cite{Giordano2025, Bartoli2025}.

\subsection{Reconfigurable Logic and FPGA-based Processors}
Ultra-Low-Power Field-Programmable Gate Arrays (FPGAs) provide custom datapaths, fine-grained precision control, and parallel sensor-fusion pipelines, trading flexibility for modestly higher static power.
The Lattice iCE40 UltraPlus \cite{iCE40UltraPlus} offers DSP blocks and 1.1 Mbit SRAM, enabling always-on Binary Neural Network (BNN) class \cite{Qin2020} inference at sub-mW power \cite{Lemieux2019}.
QuickLogic EOS S3 \cite{QuickLogicEOSS3} combines an M4F with an eFPGA fabric for dynamic workload partitioning.
The Microchip PolarFire \cite{MicrochipPolarFire} uses non-volatile flash fabric and Flash*Freeze standby to achieve ultra-low static power for burst-mode inference \cite{Chisciotti2022}.

\subsection{Neuromorphic Processors}
Neuromorphic chips use event-driven Spiking Neural Networks (SNNs) \cite{Eshraghian2023} with co-located memory/compute, offering orders-of-magnitude energy reductions for always-on sensing.
Intel Loihi 2 \cite{IntelLoihi2} integrates 128 neuro-cores and supports up to one million neurons within $\approx$100 mW for efficient video and audio processing \cite{Shrestha2024}, and targets a wide range of problems, including graph search, similarity search, LASSO regression, and combinatorial optimization \cite{Davies2021, Orchard2021}.
IBM TrueNorth \cite{Akopyan2015} implements 4,096 cores and one million neurons at $\approx$70 mW using a scalable neuro-synaptic mesh for a wide range of cognitive and sensory perception applications \cite{Akopyan2016}.
SynSense DYNAP-SE and Speck \cite{Richter2023} combine mixed-signal or digital SNNs with event-based sensing for sub-10 mW spiking computer vision \cite{Bonazzi2024}.

\subsection{Processing-In-Memory (PIM) and Compute-In-Memory (CIM)}
PIM/CIM architectures reduce data movement by performing MAC operations directly inside memory arrays, enabling highly parallel and energy-efficient Matrix-Vector Multiplication (MVM) for edge inference.
Analog CIM uses Resistive Random Access Memory (RRAM)/SRAM arrays to perform current-summation MACs with extreme density.
Mythic M1076 \cite{M1076MythicAMP} performs analog MVM using flash-based compute tiles, achieving tens of TOPS without Dynamic Random Access Memory (DRAM) accesses.

\subsection{In-Sensor Computing}
In-sensor processors execute inference directly in the sensor readout pipeline, eliminating high-bandwidth video transmission and reducing system power by orders of magnitude.
3D-stacked ConvNet-in-sensor architectures integrate DSP/NPU logic beneath the pixel array for near-zero-latency access.
Sony IMX500/501 \cite{IMX500} embeds a DSP and an NPU to output only metadata (\eg{}, object counts), significantly reducing host workload, as shown in \cite{Bonazzi2023}.
Samsung smart sensors \cite{SamsungSmartSensors} implement AI in the sensor for demonstration and subject tracking.
OmniVision's always-on sensors, such as \cite{OV6211}, perform presence and face detection at a power consumption of below 5 mW, waking the host only on valid events.
\section{Performance Evaluation} \label{sec:evaluation}

Performance evaluation is a critical component of this review, as architectural claims and design trade-offs become meaningful only when quantified under realistic workloads.
While peak TOPS and TOPS/W remain useful high-level indicators of computational capability, they often fail to reveal the practical bottlenecks that dominate ultra-low-power and in-sensor platforms: memory hierarchy limitations, dataflow constraints, activation sparsity, and the interplay between analog front-end processing and digital back-end computation.

To provide an empirically grounded comparison across heterogeneous architectures, we benchmark a representative always-on segmentation workload on three processors covering distinct design points: a RISC-V manycore MCU-class SoC (GAP9), an ARM Cortex-M55–based processor with a dedicated Neural Architecture Accelerator (STM32N6), and a stacked Complementary Metal-Oxide-Semiconductor (CMOS) in-sensor compute platform (Sony IMX500).
The evaluation focuses on latency, energy per inference, energy efficiency, and EDP.

\emph{\textbf{Why Segmentation Is a Stress Test for Edge AI.}}
Unlike classification or detection workloads, where feature maps quickly shrink, segmentation maintains large activation tensors, dramatically increasing memory footprint and internal data movement, as it preserves full spatial resolution throughout most of the network.
In addition, skip connections, multi-scale feature aggregation, and upsampling further amplify bandwidth pressure on constrained SRAM hierarchies.
These characteristics make segmentation particularly effective at exposing compute–memory bottlenecks and dataflow inefficiencies in ultra-low-power systems. 

PicoSAM2 \cite{Bonazzi2025} is the segmentation model used.
It is a compact, 336 million MAC U-Net derived architecture designed for edge workloads that operate on a 3$\times$96$\times$96 RGB input tensor, with a lightweight encoder–decoder pipeline tailored for low-latency inference on constrained hardware.

\begin{figure}[t!]
    \centering
    \includegraphics[width=\linewidth]{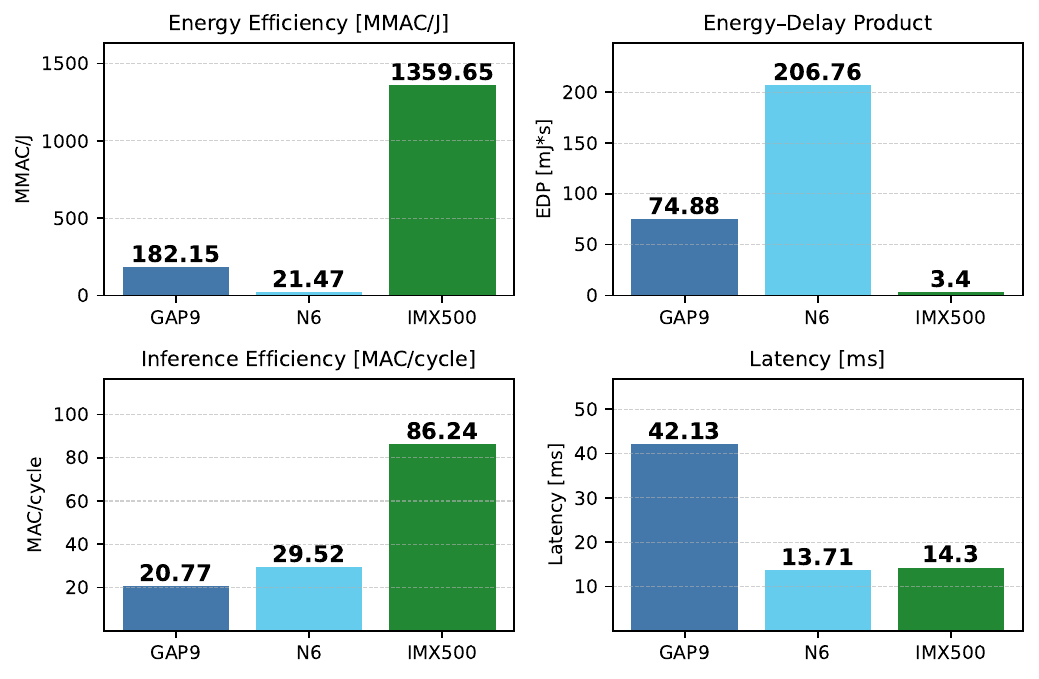}
    \caption{Benchmarking results of PicoSAM2 \cite{Bonazzi2025}, comparing its energy efficiency, latency, inference efficiency, and energy–delay product (EDP) on GAP9, STM32N6, and IMX500.
    The results highlight the advantages of in-sensor compute for improved energy efficiency and latency.}
    \vspace{-0.5cm}
    \label{fig:comparison_plots}
\end{figure}

Figure \ref{fig:comparison_plots} reports four hardware-level metrics used.
Latency quantifies end-to-end inference time, inference efficiency (MAC/cycle) measures the degree to which each architecture sustains its theoretical compute pipeline, energy efficiency (MAC/J) captures compute delivered per joule, and the energy–delay product (EDP, mJ$\cdot$s) integrates both energy cost and responsiveness into a single figure of merit.

The STM32N6 achieves the lowest latency at 13.7 ms, driven by a higher clock frequency.
The IMX500 executes the model in 14.3 ms despite a substantially lower 262.5 MHz DSP frequency, compared to the 800 MHz clock of the STM32N6.
GAP9 exhibits the highest latency (42.1 ms), which is partly attributed to its lower 370 MHz cluster clock.

The IMX500 achieves 86.2 MAC/cycle, approximately 3 times higher than the STM32N6 (29.5 MAC/cycle) and 4 times higher than the GAP9 (20.8 MAC/cycle).
This confirms that the IMX500’s tightly coupled sensor-DSP architecture offers superior spatial data reuse and reduced memory traffic.

The GAP9 achieves 182.15 MMAC/J, benefiting from low dynamic power (cluster + smart filtering unit $\approx$0.154 mW/MHz).
In contrast, the STM32N6, operating in higher power envelopes (0.8-1.2 W), yields only 21.5 MAC/J.
The IMX500 reaches 1359.6 MMAC/J, outperforming both MCU-class systems by a wide margin due to its ultra-efficient CMOS-stacked design.

The IMX500 delivers the lowest EDP (3.4 mJ$\cdot$s), indicating strong suitability for real-time mobile sensing.
The STM32N6 (206.8 mJ$\cdot$s) is dominated by power consumption, while GAP9 (74.88 mJ$\cdot$s) sits between the two, limited primarily by latency.

In summary, the IMX500 exhibits clear advantages in computing utilization, energy efficiency, and overall energy–delay balance.
GAP9 remains competitive for low-power MCU-class deployments, making it suitable for battery-constrained scenarios.
The STM32N6 offers raw speed at the cost of power, making it a suitable choice for applications where latency is a primary concern and energy efficiency is less critical.

\section{Conclusions} \label{sec:conclusions}

This work has presented an analysis of the evolving landscape of ultra-low-power Edge AI processors, focusing on architectures operating within the strict sub-200 mW power envelope required by battery-powered and energy-harvesting IoT devices.
The benchmarking of the PicoSAM2 segmentation model revealed that there is no single ``one-size-fits-all'' solution for Edge AI.
And for system designers, understanding the specific interplay between responsiveness (latency) and energy efficiency (MAC/J), or the combined balance (EDP), as revealed in this study, is essential for deploying the next generation of intelligent edge devices.

\bibliographystyle{IEEEtran}
\bibliography{bibliography}

\end{document}